\documentstyle[aps]{revtex}
\begin{document}
\author{Q.-Y. Cai}
\title{Comment on ``Quantum Dialogue''}
\address{Wuhan Institute of Physics and Mathematics, The Chinese Academy of Sciences,%
\\
Wuhan 430071,China}
\maketitle

\begin{abstract}
\begin{description}
\item  PACS: 03.67.Hk
\end{description}
\end{abstract}

Recently, Nguyen presented an entanglement-based protocol for two people to
simultaneously exchange their message [1]. The basic idea of this protocol
is beautiful. However, the detection probability of per control mode(CM) run
under disturbance attack [2] is incorrect. So the total detection
probability after N runs is not reliable. In this comment, we will show that
detection probability in every CM run under the disturbance attack is not
3/4 but 1/2.

In Nguyen's protocol [1], Bob first prepares an Einstein-Podolsky-Rosen
(EPR) that is randomly in one of the four Bell states. Bob sends one qubit
of the EPR pair to Alice and keeps another. When Alice receives Bob's travel
qubit, she performs an encoding operation on the travel and sends it back to
Bob. Bob performs an Bell basis measurement and announces his measurement
outcome. When received Bob's announcement, Alice can select to decode Bob's
information (message mode). Otherwise, she publish her encoded information
to check security of their dialogue. Let us suppose that Eve uses a
disturbance attack, i.e., she measures the travel qubit in the basis $%
B_{z}=\{|0>,|1>\}$ in every run. Suppose that Bob prepares the EPR pair in $%
|\psi ^{-}\rangle =\frac{1}{\sqrt{2}}(|0\rangle |1\rangle -|1\rangle
|0\rangle )$ and encodes ``01'' that means the state $|\psi ^{-}\rangle $
was changed into $|\psi ^{+}\rangle =\frac{1}{\sqrt{2}}(|0\rangle |1\rangle
+|1\rangle |0\rangle )$. Then Bob sends one qubit to Alice. Eve performs an $%
B_{z}$ measurement on this travel qubit and forwards it to Alice. Alice
performs an encoding operation and sends this qubit back to Bob. Bob
performs an Bell basis measurement and announces his measurement outcome.

After Eve's measurement in line $B\rightarrow A$, the state of the two
qubits is $|0\rangle _{B}|1\rangle _{A}$ or $|1\rangle _{A}|0\rangle _{B}$.
When Alice's encoding operation is $\stackrel{\symbol{94}}{1}$ or $\sigma
_{z}$, this product state does not change any more. When Alice's encoding
operation is $\sigma _{x}$ or $i\sigma _{y}$, the product state becomes $%
|0\rangle |0\rangle $ or $|1\rangle |1\rangle $. When Bob performs a Bell
basis measurement, his measurement outcome is randomly in the state $|\psi
^{-}\rangle $ or $|\psi ^{+}\rangle $. Clearly, when Alice announces her
encoding operation, Bob has a probability $p=1/2$ to find out Eve is in
line. When Alice's encoding operation is $\sigma _{x}$ or $i\sigma _{y}$,
the product state becomes $|0\rangle |0\rangle $ or $|1\rangle |1\rangle $.
Bob's final measurement out is $|\phi ^{\pm }\rangle =\frac{1}{\sqrt{2}}%
(|0\rangle |0\rangle \pm |1\rangle |1\rangle )$. The detection probability
is obviously $p=1/2$. The same conclusion can also be drawn when in other
conditions\footnote{%
That means Bob sends different states to Alice.} when Eve uses this
disturbance attack.

In this comment, we want to emphasize that when the state of the two qubit
is $|0\rangle _{B}|1\rangle _{A}$ or $|1\rangle _{A}|0\rangle _{B}$, Bob's
Bell basis measurement outcome can not be $|\phi ^{\pm }\rangle $. Also,
when the product state is $|0\rangle |0\rangle $ or $|1\rangle |1\rangle $,
Bob's measurement outcome can not be $|\psi ^{\pm }\rangle $. That is the
reason why detection probability is not $3/4$ but $1/2$ under Eve's
disturbance attack.

\section{references:}

1. B. A. Nguyen, Phys. Lett. A (in press); quant-ph/0406130.

2. Q.-Y. Cai, Phys. Rev. Lett. 91, 109801(2003).

\end{document}